\documentclass[%
 reprint,
 amsmath,amssymb,
 aps,
]{revtex4-2}
\usepackage{xcolor}
\usepackage{graphicx}
\usepackage{dcolumn}
\usepackage{bm}
\usepackage{hyperref}

\newcommand{\black}[1]{\textcolor{black}{#1}}

\begin{document}

\preprint{APS/123-QED}

\title{Metacrystals: Inversely-designed 3D-printed intelligent panels for 6G communications}

\author{Mohammad M. Asgari\textsuperscript{1}}
\thanks{Email: mohammadmahdi.asgari@aalto.fi}
\author{Peter B. Catrysse\textsuperscript{2}}
\author{\black{Shuai S. A. Yuan\textsuperscript{1}}}
\author{Haiwen Wang\textsuperscript{3}}
\author{Shanhui Fan\textsuperscript{2,3}}
\author{Viktar Asadchy\textsuperscript{1}}
\thanks{Email: viktar.asadchy@aalto.fi}
\affiliation{\textsuperscript{1}Aalto University, Department of Electronics and Nanoengineering, Espoo 02150, Finland}
\affiliation{\textsuperscript{2}Stanford University, E. L. Ginzton Laboratory, Department of Electrical Engineering, Stanford, CA 94305, USA}
\affiliation{\textsuperscript{3}Stanford University, E. L. Ginzton Laboratory, Department of Applied Physics, Stanford, CA 94305, USA}

\begin{abstract}
\noindent\textbf{Abstract:}
Metasurfaces represent a promising platform for improving coverage in future communication systems. Passive designs are especially attractive because they need no power supply and can be manufactured at low cost. However, most passive metasurfaces work well only for one polarization, frequency band, or incidence angle, which limits their practical use. Here we propose passive intelligent panels, termed metacrystals, that overcome these limitations by enabling highly complex multiplexed responses to multiple incident waves simultaneously and independently. This capability is enabled by a compact volumetric architecture that goes beyond conventional metasurfaces by exploiting a finite, yet still modest, thickness to unlock substantially more degrees of freedom. Through simulations and experiments, we demonstrate all-dielectric metacrystals capable of simultaneously controlling anomalous reflection and absorption, both in transmission and reflection regimes. Designed using inverse topology optimization, these metacrystals combine structural integrity, straightforward scalability, and compatibility with low-cost 3D printing for operation up to 100 GHz.

\begin{description}
\item[Keywords]
Metasurfaces, intelligent surfaces, smart skins,  multifunctional response, anomalous reflectors, inverse design, 6G communications, 3D printing.
\end{description}
\end{abstract}

\maketitle


\section*{\label{sec:level1}Introduction}

The advent of sixth-generation (6G) and future wireless technologies will transform communications by offering higher data rates, improved energy efficiency, and lower latency \cite{chowdhury20206g}. However, the realization of high data rates necessitates the exploration of new frequency bands, such as millimeter (mm) waves and sub-THz bands\cite{wang2018millimeter,akyildiz2014terahertz}. While these frequencies offer vast amounts of bandwidth, they also present considerable challenges due to their high atmospheric attenuation, free-space path loss, and harsher scattering effects when encountering obstacles\cite{rappaport2015millimeter}.
Therefore, reliance on traditional multipath propagation is no longer feasible, and directional beams must be used for communication \cite{chowdhury20206g}. Moreover, higher-frequency signals are often blocked by obstacles such as walls, requiring a denser network of base stations and relays.
Recently, metasurfaces, also referred to as intelligent surfaces, have been proposed to mitigate these challenges by efficiently redirecting communication signals in free space to bypass obstacles~\cite{kaina2014shaping,basar2019wireless,alexandropoulos2020reconfigurable,hassouna2023survey}. These artificial surfaces, strategically positioned on walls, ceilings, and even windows, can substantially enhance both indoor and outdoor signal coverage through anomalous reflection or refraction \cite{huawei}, requiring minimal to no energy for their operation. 

Most of the existing studies on intelligent surfaces focused on achieving reconfigurable responses \cite{hassouna2023survey}. \textcolor{black}{Programmable metasurfaces are capable of dynamically manipulating several wave characteristics, including wave vector, polarization, frequency, and wavefront, within a unified structure \cite{zhang2021wireless,wu2020space}.} However, they have proven to be too expensive for widespread adoption in the communication industry. This is primarily due to their requirement to operate at high frequencies (above 30–50 GHz), their large physical footprint (approximately one square meter) even for incorporating a single communication channel, and the need for highly tunable constituent elements~\cite{huawei}. Consequently, their non-reconfigurable (completely passive) counterparts have recently gained great attention due to their significantly lower manufacturing and maintenance costs \cite{huawei}. In fact, in many real-world scenarios, reconfigurability is not necessary because the positions of the receivers and transmitters are fixed or weakly varying. For instance, in industrial settings, machinery and sensors are usually installed in fixed locations; the infrastructure and major pathways in large public hubs remain constant; and in office environments, the locations of access points are typically fixed.

While various pathways for the analytic design of passive intelligent surfaces were proposed (e.g., anomalous reflectors~\cite{diaz2017generalized,PhysRevX.8.011036,liu2023reflectarrays}, smart skins\cite{oliveri2021holographic}, metagratings \cite{ra2017metagratings,rabinovich2018analytical,popov2018controlling}, and aperiodic gratings~\cite{li2023all}), all of them lack the sufficient versatility for realistic applications.
Indeed, in most practical scenarios, it is necessary for the intelligent surface to operate effectively across both signal polarizations, multiple frequency bands,  various angles of arrival, and even all at once. 
Realizing such versatile surfaces with current analytical or semi-analytical design techniques remains very challenging, as these techniques rely on specific homogenization models (e.g., based on polarizability~\cite{asadchy2016perfect}, susceptibility~\cite{lavigne2018susceptibility}, or surface impedance tensors~\cite{8358753}).
Factors such as frequency dispersion, nonlocality, and anisotropy make the implementation of the unit cells with required material parameters hardly possible. 
\black{Recent work on multifunctional metasurfaces at microwave and sub-THz frequencies falls into two main classes: multi-incidence and multi-dimensional. Multi-incidence designs operate under multiple incidence angles or wave vectors; examples include angle-dependent/independent metasurfaces \cite{li2025angle,JiahuiTAP2025,zografopoulos2021all,yucel2025angle}, directional Janus metasurfaces \cite{chen2020directional}, and schemes multiplexing guided and space waves \cite{guan2025guided}. Multi-dimensional designs simultaneously control several wave properties (polarization $p$, propagation direction/wavefront angle $\theta$, phase $\phi$, and amplitude $A$) typically for a single incident wave. Demonstrations include concurrent control of polarization and direction \cite{elad2024simultaneous,jing2019achieving,kazemi2020simultaneous,lu2025independent,yepes2021perfect}, wave-vector modulation across frequencies using multi-band metasurfaces \cite{xu2020wavevector,shang2022design,liu2020dual,huang2023conformal}, and co-modulation of polarization and wavefront \cite{yuan2020independent,shi2019generation}. These passive metasurfaces demonstrated to date still realize only one or two such multi-incidence/-dimensional functions at a time.} 

In this paper, we introduce the concept of metacrystals, which are all-dielectric binarized composites capable of performing multidimensional functionalities for multiple predefined incident waves concurrently and independently. 
This versatile responses are achieved by transitioning from the traditional sub-wavelength single-layer metasurface design to a bulk multilayer topology with a larger number of degrees of freedom. We use the terminology metacrystals due to their similarity to both photonic crystals (supporting multiple diffraction orders) and metamaterials (with deeply sub-wavelength building blocks).  To design the metacrystals, we employ an inverse design method using adjoint-based topology optimization~\cite{jin2020inverse} that was recently used for applications in fluid mechanics~\cite{dilgen2018topology,bendsoe2013topology}, material science~\cite{bosch2024topological}, acoustics~\cite{christiansen2019topological} and optics~\cite{jensen2011topology,piggott2015inverse,sell2017large,hughes2018adjoint,pestourie2018inverse,mohammadi2019inverse,camayd2020multifunctional, augenstein2020inverse,jin2020inverse,christiansen2021inverse,zhao2021perfect,wang2022design,catrysse2022Bayer,roberts20233d}.
This design approach provides a direct mapping from the desired complex electromagnetic response to the material topology, bypassing the need for homogenization models. 
As a proof of concept for 6G communications, we designed several metacrystals engineered to perform various functionalities for different angles of arrival and two orthogonal polarizations within a single geometry (up to six incidences). We also demonstrate the possibility of extended bandwidth (multi-frequency) response and simultaneous operation in both reflection and transmission regimes.
The designed metacrystals represent all-dielectric binarized composites with robust structural integrity, consisting of regions with low-permittivity material and air gaps. This configuration allows for cost-effective and large-scale fabrication using conventional 3D printing technology~\cite{ataloglou2023synthesis,ng2014direct,dong2019wideband}, supporting high operational frequency ranges up to 100 GHz. 
Owing to their low cost and relatively thin structure, these passive intelligent panels can be installed on building walls and ceilings in various 6G communication scenarios. To demonstrate their feasibility, we fabricated and experimentally characterized one of the designed metacrystals \black{and demonstrate its application potential for non-line-of-sight communication links.}
\begin{figure}[bt]
\centering
 \includegraphics[width=0.96\columnwidth]{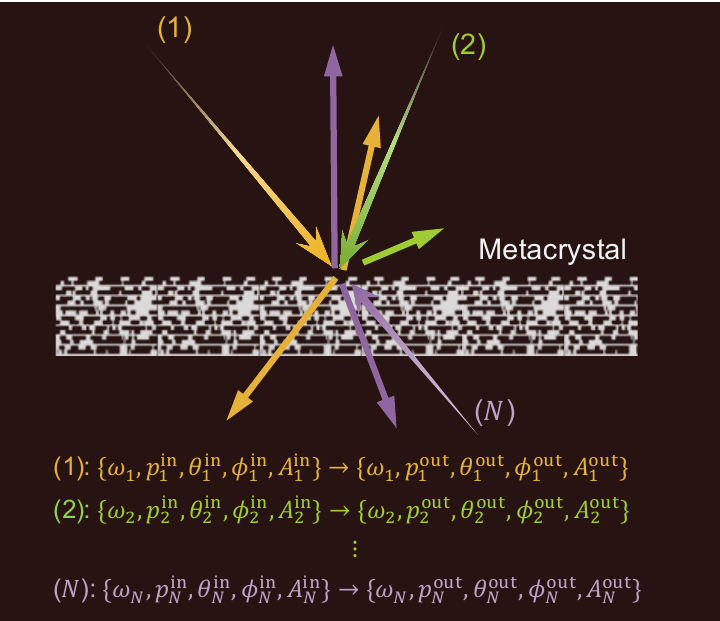}
 \caption{\textbf{Concept of a multidimensional metacrystal.} 
Schematic illustration of a metacrystal capable of independent multidimensional manipulation of multiple predefined incident waves enumerated from (1) to (N). For each incident wave, the metacrystal transforms the input parameters (polarization $p$, propagation direction $\theta$, phase $\phi$, amplitude $A$, and frequency $\omega$) into independently controlled output parameters. The volumetric patterned dielectric structure enables simultaneous control over multiple electromagnetic wave properties for distinct incidences. Symbols: $\omega$, frequency; $\theta$, propagation direction; $\phi$, phase; $A$, amplitude; $p$, polarization state.}

 \label{fig:fig0}
\end{figure}

\begin{figure*}[tb]
\centering
  \includegraphics[width=0.9\textwidth]{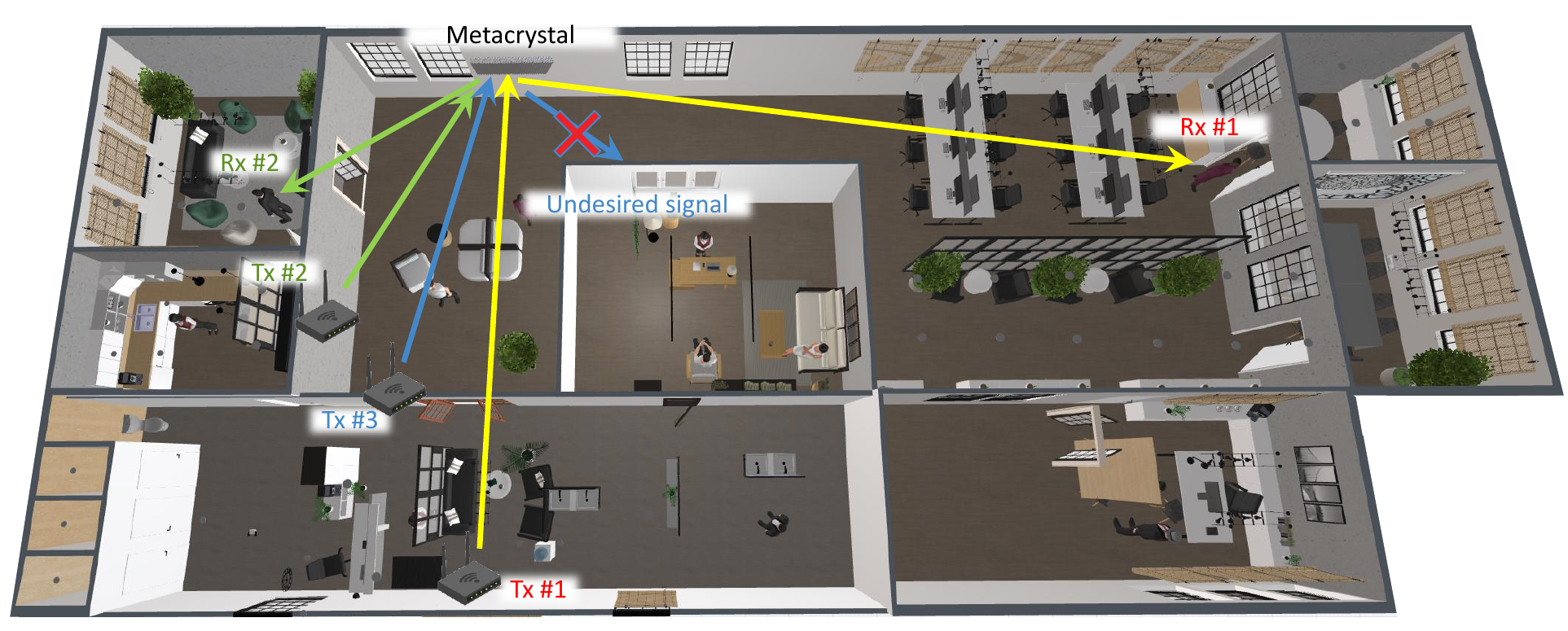}
\caption{\textbf{Passive intelligent panel deployment in an indoor wireless scenario.}
  Illustration of a metacrystal functioning as a passive intelligent panel to optimize signal routing in an office environment. The metacrystal redirects electromagnetic waves from transmitters (Tx) toward designated receivers (Rx) while suppressing undesired signal paths. Multiple independent communication links are simultaneously controlled, enabling targeted signal delivery and interference mitigation.}

  \label{fig:fig1}
\end{figure*}

\subsection*{Metacrystals for Wireless Communication}
Passive intelligent surfaces in most realistic settings need to support multifunctional responses, providing desired and generally independent operations for several channels of incident signals. Indeed, wireless communication systems predominantly operate across a range of different frequency bands with finite bandwidths, necessitating network support for these bands. Signals may originate from various transmitters and have random or specific polarizations and directions of arrival. With our passive intelligent metacrystals, we aim to implement such multidimensional multi-incidence responses into a single material platform. 
\black{We illustrate in Fig.~\ref{fig:fig0} the generic functionality of the proposed metacrystals. 
Owing to their bulk thickness and the associated large number of degrees of freedom, they are, in principle, capable of transforming $N$ incident waves with input parameters 
$\{ \omega_j, p_j^{\rm in}, \theta_j^{\rm in}, l_j^{\rm in}, \phi_j^{\rm in}, A_j^{\rm in} \}$ 
into output waves characterized by a different predefined sets of parameters 
$\{ \omega_j, p_j^{\rm out}, \theta_j^{\rm out}, l_j^{\rm out}, \phi_j^{\rm out}, A_j^{\rm out} \}$, 
where $j$ is an integer ranging from 1 to $N$. 
In practice, however, due to the finite thickness of the metacrystal, the limited resolution of its topology, and computational constraints in inverse design optimization, 
the number $N$ cannot be arbitrarily large, and the efficiency of the corresponding functionalities becomes limited.
} 

Figure~\ref{fig:fig1} depicts one possible scenario of the implementation of the  metacrystals for indoor communication. Naturally, the concept can also be equally useful for outdoor communications. In the considered scenario of office space, there are three transmitters (e.g., Wi-Fi routers) and 2 receivers (e.g., mobile phones) that need to communicate. The signal from transmitter Tx \#1 is set to be redirected by the metacrystal toward receiver Rx \#1. At the same time, the signal from Tx \#2, positioned at a different location from Tx \#1 and generally operating at a different frequency band, must be efficiently reflected toward Rx \#2. Moreover, we aim that for Tx \#3, the metacrystal behaves as an ideal absorber, preventing any parasitic reflections, especially toward the meeting space in the middle of Fig.~\ref{fig:fig1}.  
Although traditional design approach would require three separate intelligent surfaces to cover the specified functionalities, the proposed metacrystal can replace them all, saving the deployment footprint, minimizing the material use, and avoiding possible interference problems. 

\begin{figure*}[tb]
 \centering
\includegraphics[width=0.7\textwidth]{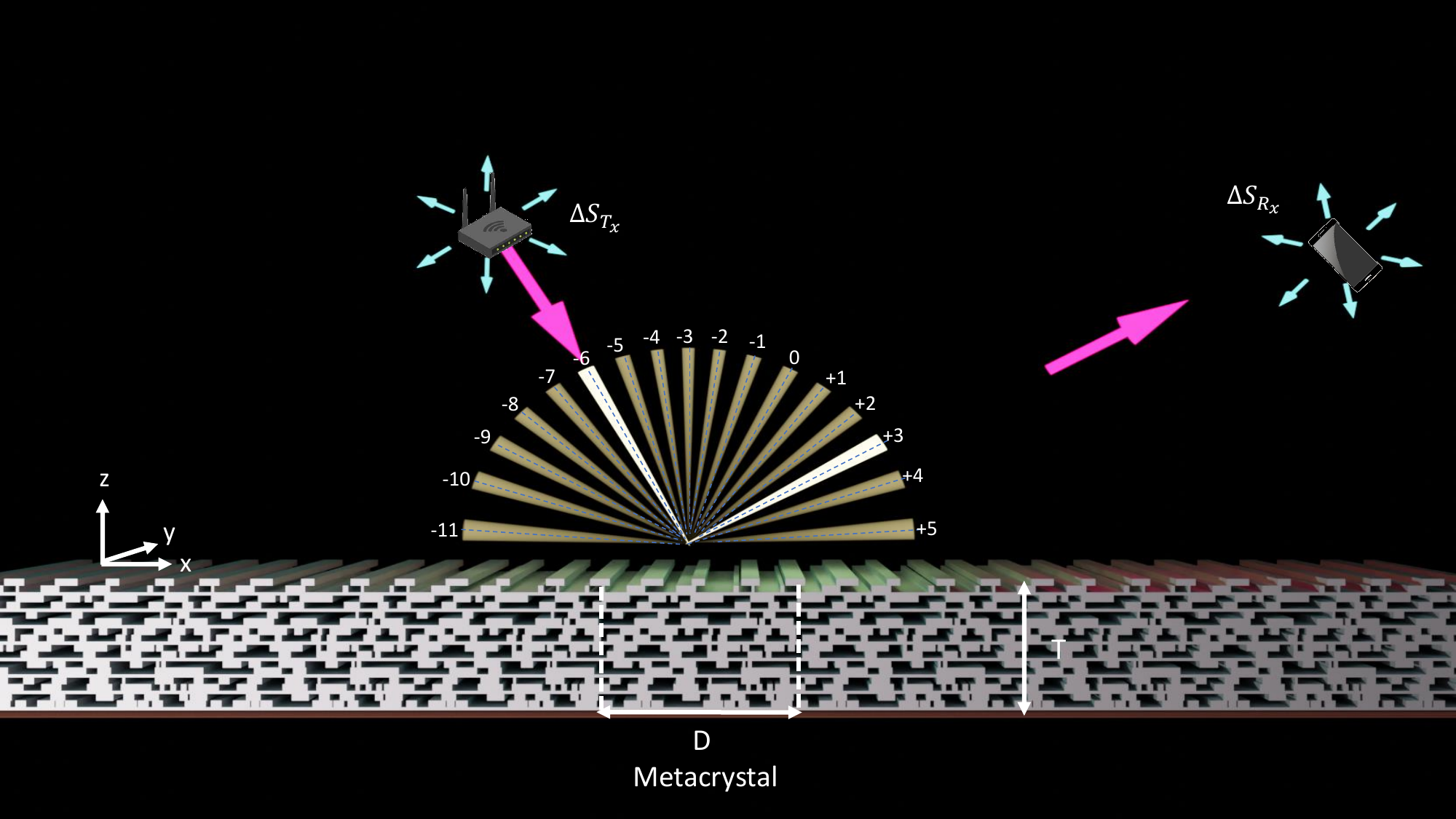}
  \caption{\textbf{Diffraction-based signal routing mechanism of the metacrystal.} 
  Illustration of a metacrystal with one-dimensional periodicity $D$ and thickness $T$. For simplicity, only one incidence is shown. The structure features an optimized spatially non-uniform permittivity distribution in the $xz$-plane and uniform geometry along the $y$-direction. The periodicity is sufficiently large to support multiple diffraction orders $n$, which discretize the angular spectrum and enable routing of signals to nearly arbitrary receiver locations. In this example, the incident signal is reflected into the $n=3$ diffraction order ($n$, diffraction order index) corresponding to the receiver direction. Owing to translational symmetry, operation remains robust against spatial displacements of the transmitter $\Delta S_{\mathrm{Tx}}$ and receiver $\Delta S_{\mathrm{Rx}}$ (cyan arrows). Yellow cones represent finite-width diffracted beams arising from the finite spatial extent of the metacrystal. The power carried by each beam is determined by the unit-cell design. The rendered structure corresponds to the fabricated metacrystal characterized in Fig.~6b. }

  \label{fig2:fig2}
\end{figure*}

\textcolor{black}{
In what follows, we assume without loss of generality that all signals (incident and reflected) propagate within the same incidence plane.
Since polarization conversion is generally undesirable in most wireless communication scenarios, leading to cross-talk between channels, we focus on metacrystal geometries with one-dimensional periodicity along the $x$-axis, with period~$D$, as illustrated in Fig.~\ref{fig2:fig2}. The metacrystal exhibits spatial variation in two dimensions (the $xz$-plane) while remaining uniform along the $y$-axis (the out-of-plane direction in Fig.~\ref{fig:fig1}). This geometry possesses mirror symmetry with respect to the $y$-axis, ensuring that the two polarization channels, transverse electric and transverse magnetic, remain orthogonal and uncoupled~\cite{meade2008photonic}. As a result, any undesired cross-talk in communications between independent polarization channels is inherently suppressed by design.}

The periodicity along the $x$-direction plays a double role. On the one hand, due to the discrete translational symmetry of the metacrystal, its operation is relatively robust to possible displacements in space of the transmitters $\Delta S_{\rm Tx}$  and/or receivers $\Delta S_{\rm Rx}$, as shown in Fig.~\ref{fig2:fig2} with cyan arrows (assuming their orientations do not change). In contrast, aperiodic intelligent surfaces are typically more sensitive to such displacements because they are designed for the exact locations of the receiver and transmitter~\cite{tcvetkova2016scanning,rengarajan2010scanning,oliveri2021holographic}. Thus, metacrystals, despite their non-reconfigurable nature, can accommodate certain variations in the wireless communication environment. 
On the other hand, the metacrystal’s periodicity allows us to significantly reduce the simulation domain for accelerating the optimization. Furthermore, with periodic structures, it is easier to tile them and align to cover larger surface areas as required. Furthermore, we note that the present work focuses on planar metacrystal topologies, which represent one of the most practically relevant deployment scenarios (e.g., panels mounted on walls and ceilings). The same adjoint-based inverse-design method can, in principle, be extended to address non-periodic or curved deployments.

Beyond the specific scenario illustrated above, the proposed metacrystal platform aligns well with several core objectives of future 6G wireless networks. These include the deployment of intelligent environments that support high user densities, angular and spectral multiplexing, and joint communication and sensing \cite{8869705}. The demonstrated ability to independently manipulate wavefronts under multiple excitation configurations enables use cases such as user-specific beam routing, interference isolation, and adaptive coverage shaping, especially in non-line-of-sight (NLoS) or reflection-dominated environments. Moreover, the passive, fabrication-friendly nature of the metacrystal makes it an attractive candidate for static infrastructure integration, where low cost, low power, and high directional control are prioritized. In contrast to active reconfigurable metasurfaces that require external control circuitry, our approach enables high-efficiency, multidimensional signal management in a fully passive format, suggesting a pathway toward metasurface-enabled wireless systems where compact, multifunctional diffractive hardware plays a fundamental role.

\begin{figure*}[tb]
\centering
  \includegraphics[width=0.9\textwidth]{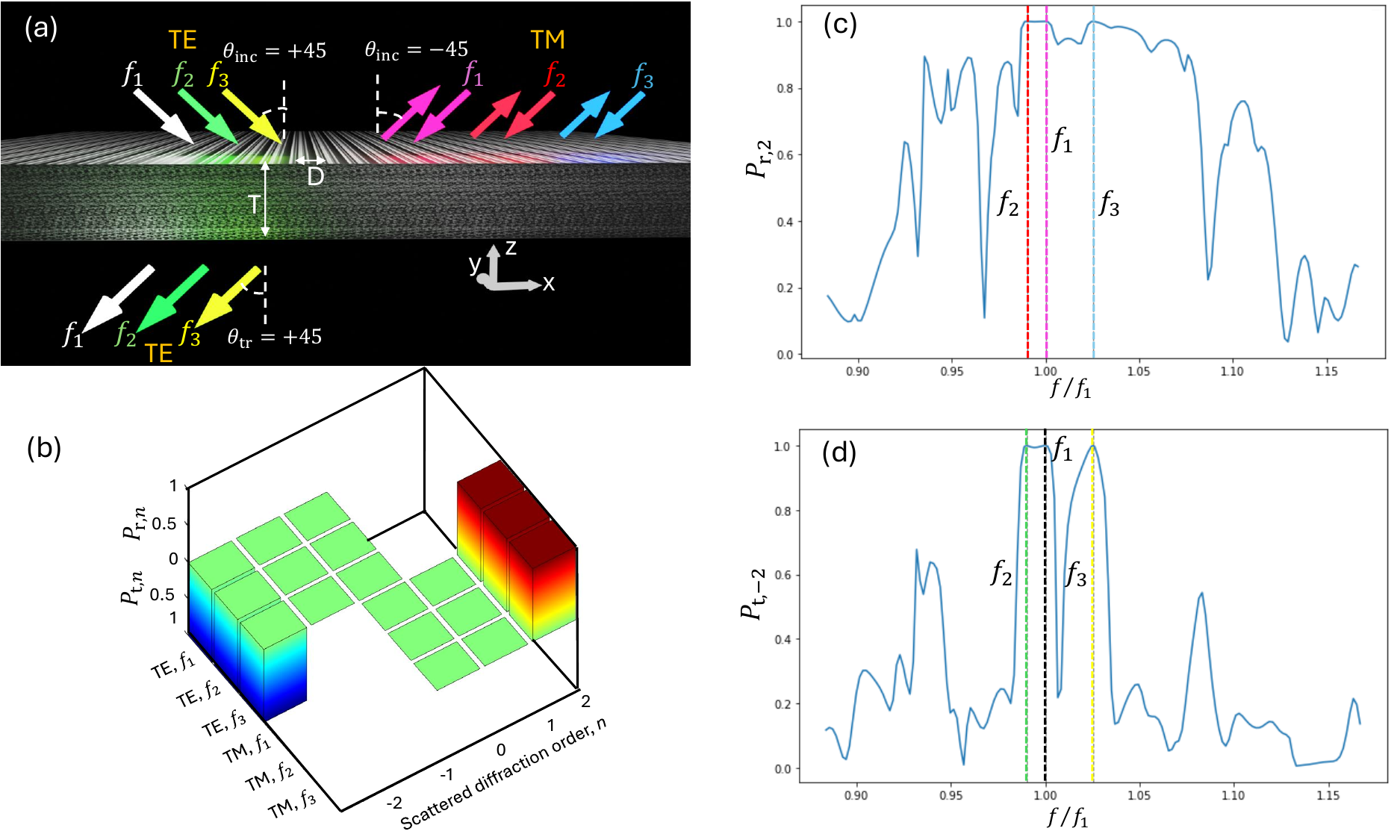}
 \caption{\textbf{Multiband multifunctional response of the first metacrystal demonstrator.}
(a) The optimized topology of the first demonstrator metacrystal providing simultaneously 6 functionalities for TE and TM  waves over extended bandwidth (frequencies $f_1$, $f_2$, and $f_3$) incident at $+45^\circ$,  and $-45^\circ$ from the normal. While the TE waves experience perfect negative refraction, the TM waves are fully reflected in the retroreflection direction. The metacrystal is uniform along the $y$-direction. Arrows of different colors denote different illumination scenarios.  
  (b) The simulated distribution of the reflected and transmitted power into different diffraction channels for the 6 incidence scenarios. The powers are normalized by the incident power and by the ratio of the wave impedances. The upward-oriented color columns denote the reflected powers and the downward-oriented columns stand for the transmitted powers. The blank spaces in the plot correspond to closed (evanescent) diffraction orders. The reflection/refraction angle of the $n$-th diffraction order can be determined according to $\theta_{n}= \arcsin (\sin \theta_{\rm inc}+ n \lambda/D)$.
  (c) Efficiency of the retroreflection for the TM waves (normalized reflected power into the $n=+2$ diffraction order) versus the incident frequency. Vertical dashed lines indicate the design frequencies $f_1$, $f_2$, and $f_3$. The operational bandwidth is strongly increased due to the close spacing of the designed frequencies $f_1$, $f_2$, and $f_3$. (d) Efficiency of the negative refraction for the TE waves (normalized transmitted power into the $n=-2$ diffraction order) versus the incident frequency. The bandwidth is partially increased. Further increase can be achieved by adding an additional frequency point between $f_1$ and $f_3$.} 

 \label{fig:final_dem_1}
\end{figure*}

\subsection*{Demonstrators}
In this section, we aim to showcase three demonstrator examples of metacrystals. 
\textcolor{black}{For the first two demonstrators, we present the optimized designs using grayscale permittivity distributions to illustrate the concept with concrete metacrystal examples. We note that the use of gray scale permittivity can in fact be achieved in practice and has been demonstrated at microwave frequencies \cite{huang2023conformal}. For the third demonstrator, which we fabricate using 3D printing and characterize experimentally, we additionally introduce thin supporting layers to ensure structural integrity and apply permittivity binarization to make it suitable for implementation with our 3D printing capabilities.
Given that this work targets metacrystals for wireless communication applications, the demonstrators focus on the most relevant functionalities, including control of frequency, polarization, and angle of incidence. We aim to achieve multidimensional control of wave amplitude and propagation direction in both reflection and transmission regimes for both polarizations.
}

For the first demonstrator, we target a highly-complex manipulation of multiple incident waves with extended bandwidth (3 different frequencies $f_1$, $f_2$, and $f_3$), 2 different polarizations, and 2 angles of incidence, as depicted in Fig.~\ref{fig:final_dem_1}(a). Furthermore, the metacrystal operates both in transmission and reflection regimes simultaneously.  
It exhibits strong asymmetry in both in-plane and out-of-plane wave manipulation: transverse magnetic (TM-polarized) waves at the three different frequencies, incident at $\theta_{\rm inc} = -45^\circ$, undergo complete retroreflection, akin to negative reflection from an opaque medium. Conversely, transverse electric (TE-polarized) waves at the same frequencies, incident at $\theta_{\rm inc} = +45^\circ$, experience full negative refraction (transmission), as if the structure were a fully transparent Veselago medium slab.

The metacrystal is designed with a periodicity of \( D = \sqrt{2}\lambda \).
Here, we assume the grayscale lossless
permittivity distribution with $\varepsilon_{\rm max} = 5$, which is a realistic value for low-loss dielectrics in the 
mm-wave frequency band. These parameters enable efficient diffractive behavior while maintaining subwavelength features in each layer.  While the metacrystal can be designed to operate at widely separated frequencies, here we deliberately selected three closely spaced frequencies $f_1=100$~GHz, $f_2=99$~GHz, and $f_3=102.53$~GHz. With this choice, we investigate the capability of extending the operational bandwidth of the metacrystals, a highly desirable feature for applications in wireless communications. 
Under such configuration, for the incident waves at $\theta_{\rm inc} = \pm 45^\circ$, there are 6 open diffraction orders (3 in reflection and 3 in transmission).

For the optimization, we discretize the metacrystal unit cell into a finite amount of voxels in the $xz$-plane and adjust their permittivities to achieve the desired multifunctional response. For the first demonstrator,  we choose the number of voxels in the $z$- and $x$-directions to be 120 and 150, respectively, to provide sufficient degrees of freedom for multifunctional control while maintaining compatibility with fabrication and computational constraints. The voxel dimensions are $t = 0.047\lambda$  and $d = 0.0094\lambda$.
To understand how different metacrystal parameters influence the optimization of its response, we refer the reader to 
Supplementary Section 1 where we performed a corresponding systematic study. Although the total thickness  $T=5.6\lambda$ is large compared to the wavelength, its absolute value is only 16.8~mm for an operational frequency of 100~GHz. Such metacrystals could be readily deployed on internal or external walls of buildings, appearing as nothing more than futuristic passive thin panels. The permittivity distribution of the metacrystal at all voxels is detailed in the Supplementary Data 1.

The power distribution across the reflected and transmitted diffraction channels for the optimized metacrystal is calculated using rigorous coupled wave analysis (RCWA) and plotted in Fig.~\ref{fig:final_dem_1}(b).
The plot depicts the reflected \( P_{{\rm r},n} \) and transmitted \( P_{{\rm t},n} \) powers normalized to the incident power and the impedance contrast~\cite{asadchy2017flat} for each frequency-polarization pair. A clear pattern is evident with the
near-unity-amplitude columns, signifying the efficient manipulation of six incident waves into the designated reflected and transmitted diffraction channels. The average efficiency, calculated as
the arithmetic average of the efficiencies for each incident wave scenario, reaches nearly 99.99\%. 
Figures~\ref{fig:final_dem_1}(c) and (d) depict the efficiencies of the retroreflection for the TM waves and negative refraction for the TE waves, respectively, versus frequency. One can observe the efficiencies reaching nearly unity at all three frequencies $f_1$, $f_2$, and $f_3$ chosen for extended bandwidth operation. Remarkably, due to the deliberate proximity of the three frequencies, we could obtain half-power bandwidth of 2.54 \% and 11.01 \% for the TM and TE waves, respectively. For comparison, when we optimize metacrystal with the same functionalities but for a single $f_1$ frequency, the bandwidths are much smaller: 0.9\% and 1.09\% for the TM and TE waves, respectively. This performance implies that further bandwidth increase of the metacrystal can be achieved through the concurrent optimization of additional frequency points. 

\textcolor{black}{To highlight the significant advantage of bulk metacrystals over two-dimensional metasurfaces for achieving multifunctional, multi-angle responses, we optimized for the same functionality as shown in Fig.~\ref{fig:final_dem_1}(a), but using a single-layer topology. This ``metasurface-like'' design employed the same voxel density in the $x$-direction and a typical metasurface thickness of $t = T = 0.25\lambda$. In this case, the average efficiency across the six functionalities did not exceed 2.28\%. This result underscores the critical role of metacrystal thickness in enabling high-performance multifunctional responses.
}

To further demonstrate the versatility of our metacrystal design framework, we present a second demonstrator targeting independent multifunctional control under various incident angles and distinct polarization states.  This scenario highlights the ability of the metacrystal to simultaneously accommodate independent wavefront transformations, extending the concept to the angular and polarization domains. 
Here, we concentrate on the scenario of a reflective metacrystal since intelligent surfaces are typically deployed on top of opaque walls of buildings, preventing signal transmission.
In order to effectively discretize the angular space in the $xz$-plane, we need to allow a sufficiently large number of diffraction orders. Without the loss of generality, we choose a periodicity of $D=4.2\lambda$, which enables us to achieve 9 diffraction channels for normal incidence or 8 and 9 for the considered oblique illuminations. This discretization of angular space is essential for realizing multifunctional control of waves. A smaller period would reduce the number of available diffraction orders as well as the degrees of freedom, limiting control over wavefronts and angular resolution~\cite{wang2022design}. Conversely, a much larger period would increase the size and thickness of the metacrystal, making it less compact and more prone to undesired scattering.
We found that $D = 4.2\lambda$ represents a balanced trade-off: small enough to maintain a compact footprint but large enough to open multiple diffraction orders for angular control. 

\begin{figure}
  \centering
\includegraphics[width=0.45\textwidth]{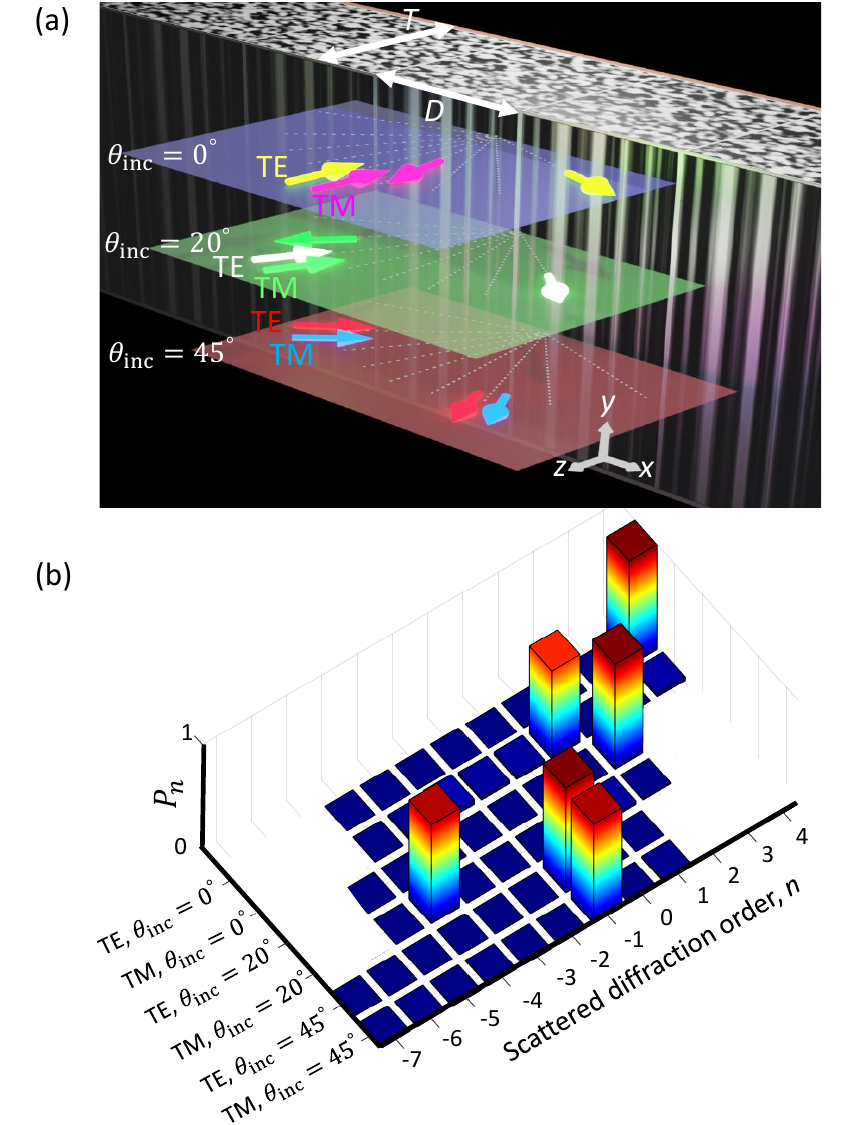}
  \caption{\textbf{Angular multiplexing in the second metacrystal demonstrator.} (a) The optimized topology of the second demonstrator metacrystal providing simultaneously 6 functionalities for TE and TM  signals incident at $0^\circ$, $20^\circ$, and $45^\circ$. For visual clarity, each pair of functionalities is depicted in a separate incidence plane (the metacrystal is uniform along the $y$-direction). Arrows of different colors denote different illumination scenarios. The thin dashed black lines depict the diffraction order directions for a given incidence.  
  (b) The simulated distribution of the reflected power into different diffraction channels for the 6 incidence scenarios. Due to the lossless metacrystal response, the total sum of reflected power for each incidence is equal to unity. The blank spaces in the plot correspond to closed (evanescent) diffraction orders. The reflection angle of the $n$-th diffraction order can be determined according to $\theta_{{\rm r}, n}= \arcsin (\sin \theta_{\rm inc}+ n \lambda/D)$.  }
 \label{fig:fig_gen}
\end{figure}

For this demonstrator, we require lossless operation and consider 6 functionalities chosen arbitrarily (but respecting energy conservation and reciprocity) that must be satisfied simultaneously. The number of simultaneous functionalities is not fundamentally limited. A larger number typically requires a metacrystal with larger thickness (see Supplementary Section 1.1). The permittivity of the materials is assumed to be purely real. 
The 6 functionalities are shown in Fig.~\ref{fig:fig_gen}(a) using three incidence planes (blue, green, and red) separated along the $y$-direction for visual clarity. In each plane, the desired control of waves for two polarizations (TE and TM) is shown.
We aim that for incident signals at $0^\circ$, the TE and TM polarizations are reflected into the $n=+4$ and $n=+1$ diffraction channels, respectively, as indicated with yellow and purple arrows in Fig.~\ref{fig:fig_gen}(a).  This configuration corresponds to a polarization-selective anomalous reflection splitter.
On the other hand, for an incident angle of $20^\circ$, the TE (white arrow) and TM (green arrow) signals are directed towards positive and negative diffraction channels, that is, the $n=+2$ and $n=-4$ channels, respectively. For the last two functionalities, we require polarization-independent anomalous reflection towards the $-1$-st diffraction order when TE and TM signals are incident at $45^\circ$. Importantly, the incident angles ($0^\circ$, $20^\circ$, and $45^\circ$) were chosen completely arbitrarily. This example thus illustrates that we can select the angles of arrival from different transmitters independently.

For the second demonstrator,  we choose the number of voxels in the $z$- and $x$-directions to be 70 and 150, respectively. With the voxel dimensions $t=0.168\lambda$ and $d= 0.028\lambda$, this translates into the total thickness of $T=11.76 \lambda$ and period $D=4.2\lambda$. 
Here, we assume the grayscale permittivity distribution with $\varepsilon_{\rm max}=5$.  

The optimized topology of the metacrystal can be seen in Fig.~\ref{fig:fig_gen}(a). 
The permittivity distribution of the metacrystal at all voxels is detailed in the Supplementary Data 2.

The power distribution across the reflected diffraction channels for the optimized metacrystal is plotted in Fig.~\ref{fig:fig_gen}(b). A clear pattern is evident with the near-unity-amplitude columns, signifying the efficient reflection of six incident waves into the designated reflection channels. The reflected powers for $\theta_{\rm inc}=0^\circ$ reach $P_{+4}^{\rm TE}=99.99\%$ and $P_{+1}^{\rm TM}=83.92\%$,
for $\theta_{\rm inc}=20^\circ$, we obtain $P_{+2}^{\rm TE}=99.99\%$ and $P_{-4}^{\rm TM}=94.63\%$, and for $\theta_{\rm inc}=45^\circ$, $P_{-1}^{\rm TE}=99.99\%$ and $P_{-1}^{\rm TM}=95.24\%$.
The overall efficiency for the six functionalities calculated as the arithmetic average of the reflected powers specified above reaches 95.62\%, showcasing the effectiveness and precision of the designed method. 
As is seen from Fig.~\ref{fig:fig_gen}(b), the parasitic scattering to other diffraction channels is negligible. 
Indeed, the cross-talk into non-target diffraction orders remains below 5\% for most functionalities. Specifically, for the TM mode at $\theta_{\rm inc}=0^\circ$, a higher leakage is observed, whereas in the remaining five cases, the power scattered into secondary channels is significantly lower. For example, the power in orders $P_{-2}^{\rm TM}$, $P_{-1}^{\rm TM}$, $P_{0}^{\rm TM}$, and $P_{4}^{\rm TM}$ with highest cross-talk for $\theta_{\rm inc}=0^\circ$ is measured as 4.82\%, 2.85\%, 2.83\%, and 3.47\%, respectively. These results demonstrate high channel selectivity and robust performance of the metacrystal.

\begin{figure*}[ht]
 \centering
 \includegraphics[width=0.9\textwidth]{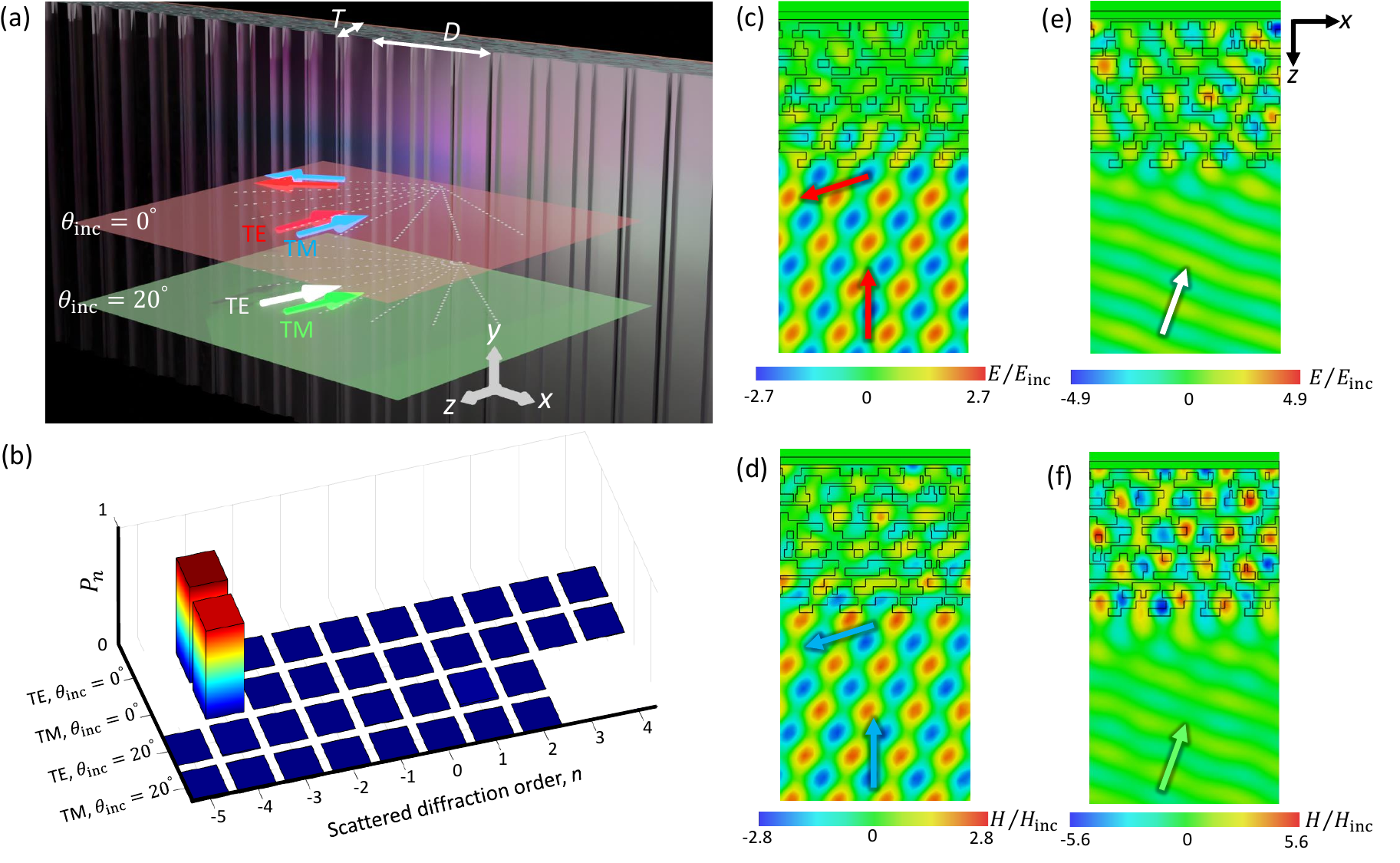}
  \caption{\textbf{Multifunctional response of the third metacrystal demonstrator.} (a) The optimized topology of the third demonstrator metacrystal providing simultaneously 4 functionalities for TE and TM  signals incident at $0^\circ$ and $20^\circ$. For visual clarity, each pair of functionalities is depicted in a separate incidence plane. Arrows of different colors denote different illumination scenarios. The thin dashed black lines depict the diffraction order directions for a given incidence.  (b) The simulated distribution of the reflected power into different diffraction channels for the 4 incidence scenarios. The blank spaces in the plot correspond to closed (evanescent) diffraction orders. For two incident scenarios, the metacrystal absorbs nearly all the incident power, as evidenced by the minimal overall reflection.
  (c)--(f) The simulated total fields (sum of incident and reflected fields) for four illumination scenarios. Fields are normalized to the incident field amplitude.  
}
 \label{fig:dem2}
\end{figure*}

Finally, for the third demonstrator, our objective is to fabricate and experimentally characterize the proposed metacrystal. To streamline the experimental verification, which involves complex angular-resolved measurements of the far fields for each incidence scenario, we limit the number of functionalities to four. The desired response of the third demonstrator metacrystal is shown in Fig.~\ref{fig:dem2}(a).  In this setup, we require polarization-insensitive response, as it is essential in many practical situations. Achieving polarization-insensitive response is non-trivial and requires comparable structural complexity as in the scenario where different polarizations experience different prescribed functionalities~\cite{hu2023cavity,shklarsh2023semianalytically}.
The TE and TM signals normally incident onto the metacrystal must be anomalously reflected into the $n=-4$ diffraction channel, corresponding to a reflected angle of 72$^\circ$. On the other hand, all signals incident at $\theta_{\rm inc}=20^\circ$ must be completely absorbed by the metacrystal. This functionality could be useful in scenarios like the one shown in Fig.~\ref{fig:fig1} to avoid possible parasitic interferences (see the blue arrows). 
To significantly simplify the fabrication process, we optimized the metacrystal using a binarized permittivity distribution (see details in the Experimental Section). This approach enables us to use only one material during fabrication, with a permittivity of $\varepsilon_{\rm max}=2.2$, which alternates spatially with air gaps having a permittivity of $\varepsilon_{\rm min}=1$. This configuration facilitates straightforward fabrication using 3D printers, particularly the most common and cost-effective filament-based printers. The mentioned permittivity value was experimentally determined for the low-loss material used in the fabrication, polyacrylic acid (UltiMaker PLA of silver color), as explained in the Experimental Section. \textcolor{black}{It should be noted that other commercially available printer filament materials can also be used. For example, filaments such as ``Zetamix $\varepsilon$'' offer low-loss permittivity values of up to 10.
} The measured loss tangent of our PLA material is $\tan\delta=0.01$.

The smallest feature size of the designed metacrystal, obtained by applying the blur function, was $0.056\lambda=0.34$~mm, which is well within the accuracy range of commercial 3D printers. To ensure structural integrity and maintain interconnection among all voxels, we added thin homogeneous layers (with $0.336\lambda$ thickness) between each two layers of optimized voxels in the $z$-direction. The optimized structure comprised only 16 voxel layers, each with a thickness of $t=0.168\lambda$. Combined with the homogeneous structural layers with $0.071\lambda$, the total thickness was $T=3.26\lambda$. Other parameters, $D$, $t$, and $d$ are the same as in the first demonstrator. 

The geometry of the optimized third demonstrator is shown in Fig.~\ref{fig:dem2}(a), with a view of the $xz$ plane provided in Fig.~\ref{fig2:fig2}. The permittivity distribution of the metacrystal at all voxels can be found in the Supplementary Data 3. The multifunctional response of the metacrystal simulated using RCWA is shown in Fig.~\ref{fig:dem2}(b).  
The designed structure exhibits anomalous reflection efficiencies of $84.39\%$ for TE polarization and $78.26\%$ for TM polarization for normally incident signals. Remarkably, despite being composed of a low-loss dielectric backed by a ground plane, the same structure nearly fully absorbs both polarizations at an incident angle of $20^\circ$, with absorption rates of $90.47\%$ and $96.34\%$ for TE and TM polarizations, respectively. 

Furthermore, in Figs.~\ref{fig:dem2}(c)--(f), we present the normalized electric field distributions for TE-polarized incidences and the magnetic field distributions for TM-polarized incidences across the four illumination scenarios. The data were obtained using CST Studio Suite. Notably, in the anomalous reflection scenarios, almost no field hot spots are generated within the optimized metacrystal, and the total field outside it exhibits a partially standing-wave pattern due to the superposition of the normally incident plane wave and the single plane wave reflected at $72^\circ$. Conversely, in the full absorption scenarios depicted in Figs.~\ref{fig:dem2}(d),(f), strong field hot spots are generated inside the crystal, leading to the high absorption. The total fields outside the metacrystal are predominantly characterized by a single incident plane wave, indicating minimal parasitic scattering into open diffraction channels. The average efficiency of the third metacrystal demonstrator reaches 80\%, which is strong given the fabrication constraints, especially considering the challenges posed by permittivity blurring and binarization due to fabrication limitations, the small overall thickness, and the presence of uniform layers for structural integrity. 


 \section{Far-Field Measurements} 
\begin{figure*}[ht]
 \centering
 \includegraphics[width=0.95\textwidth]{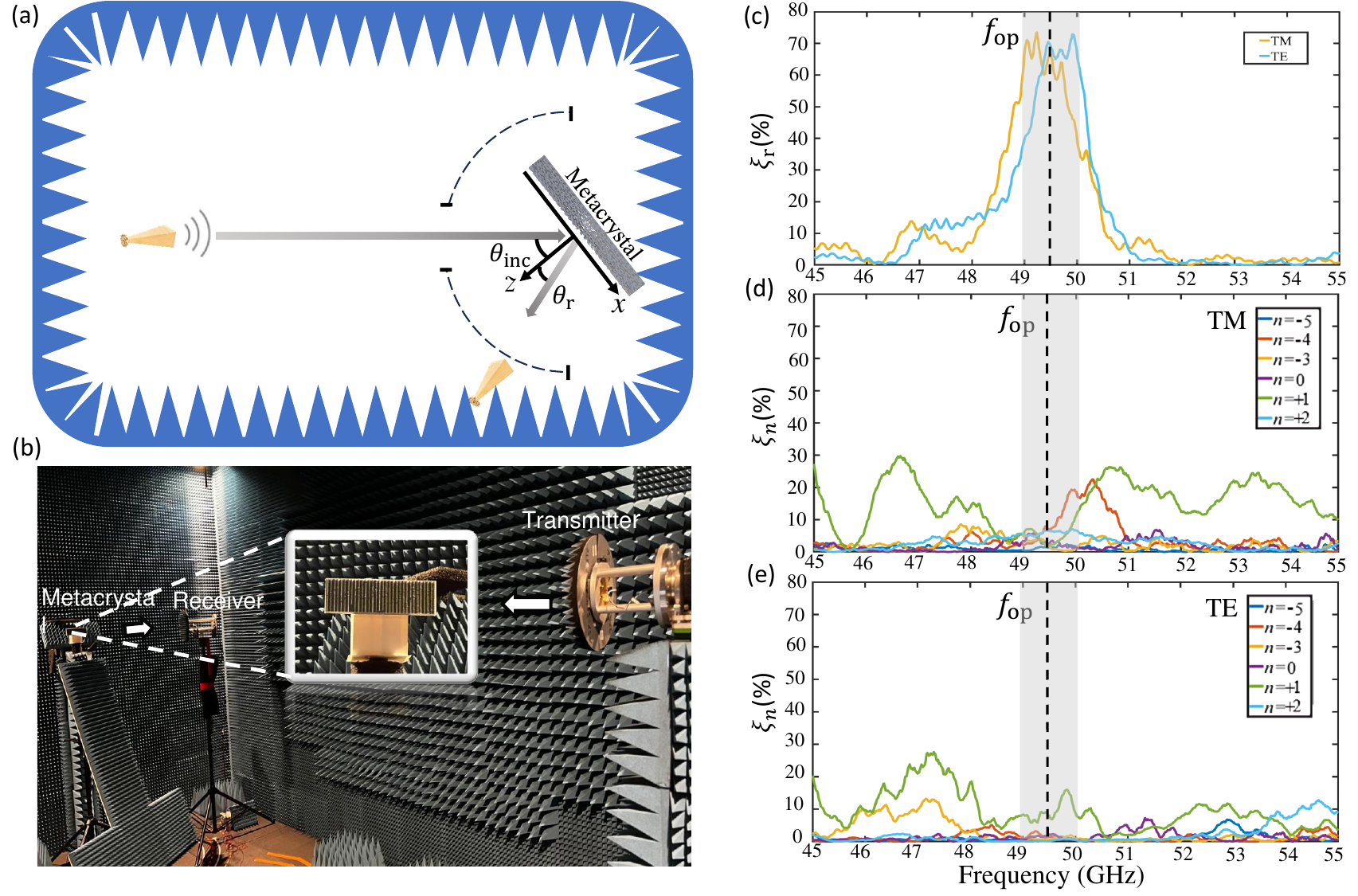}
  \caption{\textbf{Experimental validation of the metacrystal performance.} (a) The schematic of the experimental setup. (b) Experimental setup in an anechoic chamber with the inset depicting the zoomed-in image of the metacrystal sample. Several large images of the fabricated sample can be found in Supplementary Section 2. (c) The efficiency of the anomalous reflection for TE and TM polarized waves incident normally onto the sample. The dashed vertical line depicts the operational frequency $f_{\rm op}$. The grey shaded regions denote the half-power bandwidth for anomalous reflection functionality. (d) The parasitic scattering efficiency $\xi_n$  across different diffraction channels for TM polarized waves incident at an angle $\theta_{\rm inc}=20^\circ$ onto the sample. At frequency $f_{\rm op}$, the overall parasitic scattering is small, confirming the high absorption exhibited by the sample. (e) Same as in (d) for TE polarization.  } 
  \label{fig:exp}
\end{figure*}

\black{\subsection*{Measurements of the metacrystal scattering parameters}} 
The fabricated sample of the third demonstrator (see Fig.~\ref{fig:exp}(b)) comprises
8 unit cells along the $x$ axis and has the size of $33.6 \lambda = 20.16$~cm, $10.5 \lambda = 6.3$~cm, and $3.26\lambda = 1.95$~cm along the $x$, $y$, and $z$ directions, respectively.  We choose the operational frequency of the metacrystal to be 50~GHz, although with the currently used 3D printer, one can reach operational metacrystal frequencies up to 100~GHz. 
Characterizing the metacrystal at 50~GHz using quasi-optical setups, such as those in~\cite{wang2018extreme}, is particularly challenging. This is because Gaussian beams must be very wide (with a small span in momentum space) to resolve the sharp angular response of the metacrystal. Therefore, the operation of the fabricated metacrystal was
verified by measurements in an anechoic chamber emulating the free-space environment (see more details in Methods). In this case, due to the large distance between the transmitting antenna and the sample under test, the incident wavefront can be well approximated as planar. However, since the entire surface of the metacrystal is illuminated, it is crucial to account for edge scattering effects when evaluating its efficiency. To address this, we employ the measurement method proposed in~\cite{diaz2017generalized, asadchy2017flat}.

The schematic of the experimental setup can be seen in Fig.~\ref{fig:exp}(a). 
Two sets of experiments were performed to validate the demonstrator functionality for four incidences. In the first set of experiments, we measured the anomalous reflection of normally incident waves toward the $-4$-th diffraction order for both TE and TM polarizations.
The orientation and position of the platform with the metacrystal and the transmitting antenna were fixed.   The receiving antenna was moved around the metacrystal at a fixed distance from it, always facing the metacrystal (see Fig.~\ref{fig:exp}(a)). 
First, we measured the received signal at the desired $-4$-th diffraction order. For waves at 50~GHz, we observed that the maximum of the diffracted peak occurred at $\theta_{\rm r}=76^\circ$, which slightly deviates from the theoretical angle of 72$^\circ$. The observed deviation of the diffracted peak from the theoretical angle is attributed to finite-size effects (the fabricated sample comprises only eight periods), as well as fabrication imperfections such as slight variations in periodicity and surface quality introduced during 3D printing, which locally perturb the ideal periodic structure.
 Next, we measured the received signal in the reference case where the metacrystal was replaced by a metal plate of the same dimensions tilted at an angle $\theta_{\rm r}/2=38^\circ$ (to ensure a strong specularly reflected signal). Using these two measurements and the theory based on the physical optics approximation (see the Experimental Section and Supplementary Section 2), we determine the anomalous-reflection efficiency of the metacrystal for both polarizations, as plotted in Fig.~\ref{fig:exp}(c). At a frequency of $f_{\rm op}=49.475$ GHz (shown with dashed vertical line), the measured efficiency $\xi_{\rm r}$ reaches values of 70.2\% and 66.3\% for the TE and TM polarized waves, respectively, with peak values of 72.8\% and  73.4\% for TE and TM at slightly different frequencies of 49.91 GHz and 49.24 GHz, respectively. This is slightly lower than the simulated values $84.4\%$ and $78.3\%$ reported above, likely due to fabrication imperfections and the approximation of physical optics used to account for the finite size of the metacrystal. We note that these measured efficiencies should be regarded as conservative for the present small-aperture prototype (only eight periods). Increasing the effective aperture is expected to improve agreement with the simulated results because the physical-optics assumptions hold better for larger apertures. Moreover, enlarging the aperture does not necessarily increase fabrication imperfections: in practice, the area can be scaled by tiling identical metacrystal panels printed on the same system under the same process settings, so the dominant tolerance statistics (per unit area) remain essentially unchanged. Motivated by this, below we investigate a larger-aperture tiled sample and confirm its strong performance in enhancing a practical communication link.
 
 From Fig.~\ref{fig:exp}(c), one can observe the operational bandwidth of the metacrystal for anomalous reflection of approximately 2\% (shown with grey shaded region). Outside this bandwidth, the metacrystal continues to exhibit strong anomalous reflection, but the reflection angle deviates from the desired value, causing the receiving antenna, which is fixed in position, to fail to detect the reflected power. 



In the second set of experiments, we measured the absorption characteristics of the metacrystal when it was illuminated at an angle of $\theta_{\rm inc}=20^\circ$. 
To evaluate the absorption level, we measured the parasitic scattering from the metacrystal into multiple open diffraction channels. For each diffraction channel, a reference measurement was also performed using a metal plate (as described earlier) to determine the parasitic scattering efficiency, $\xi_n$, for a given channel $n$. 
The measured parasitic scattering is plotted in Figs.~\ref{fig:exp}(d) and (e) for TM and TE polarizations, respectively. In these measurements, the receiving antenna was positioned in the direction of a given diffraction channel $n$ calculated at operational frequency $f_{\rm op}$ (for each diffraction channel). This way, the absorption efficiency can be calculated as $\xi_{\rm a} = 1 - \sum_n \xi_n$, but only at frequencies close to $f_{\rm op}$. Indeed, during the frequency sweep, the positions of the diffraction channels continuously shift and do not overlap with the location of the receiving antenna.
As seen from Figs.~\ref{fig:exp}(d) and (e), for frequencies close to $f_{\rm op}$, parasitic scattering is strongly suppressed, confirming high absorption efficiency.
The absorption efficiencies  at $f_{\rm op}$ for both polarizations (dashed vertical lines in Figs.~\ref{fig:exp}(d) and (e)) reach impressive values of $\xi_{\rm a}^{\rm TE}= 87.52\%$ and $\xi_{\rm a}^{\rm TM}= 78.24\%$, respectively. 
It is worth noting that due to the physical obstruction caused by the receiving antenna, we were unable to measure parasitic scattering into two diffraction orders, $n=-1$ and $n=-2$. However, scattering into these directions is expected to be minimal according to the simulated data.
A comparison between the experimental and simulated results for the absorption scenario at the operating frequency can be found in the Supplementary Table 1.


Finally, when considering the four functionalities in the experimental setup, the efficiencies are reported as $\xi_{\rm r}^{\rm TE} = 70.2\%$, $\xi_{\rm r}^{\rm TM} = 66.3\%$, $\xi_{\rm a}^{\rm TE} = 87.5\%$, and $\xi_{\rm a}^{\rm TM} = 78.2\%$. The overall average efficiency is calculated to be 74.9\% at $f_{\rm op}$.

\begin{figure*}[ht]
\centering
 \includegraphics[width=\textwidth]{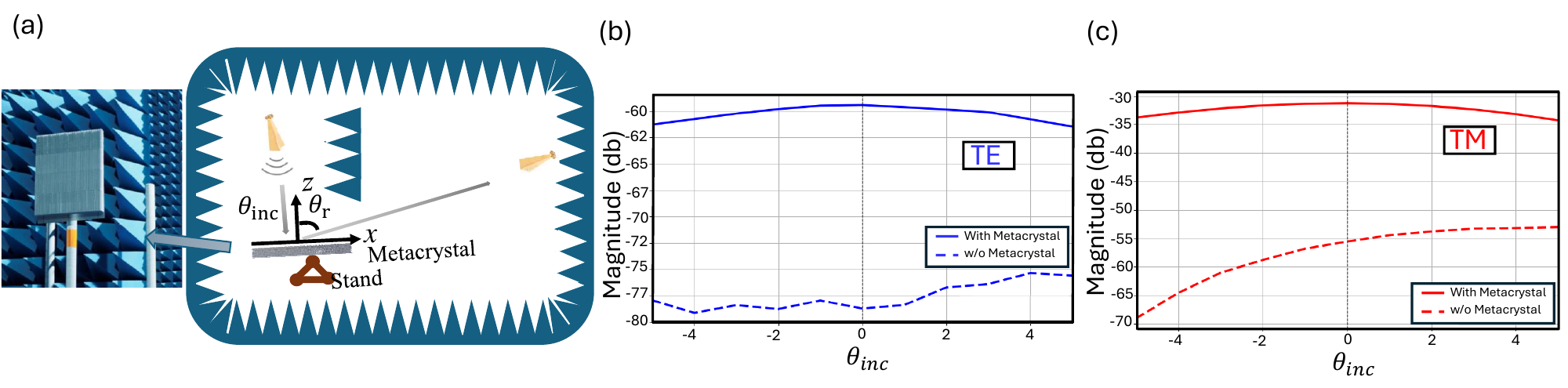}
  \caption{\textbf{Communication link measurement with the metacrystal.} \textcolor{black}{(a) Photograph and schematic of the setup used for communication link measurements. The metacrystal sample is illuminated by a TX horn antenna at a near-normal angle, and the received signal is captured by an RX horn antenna placed at the  angle $\theta_{\rm r}=72^\circ$ with respect to the $z$-axis.
(b) Measured received signal magnitude versus the incident angle for TE  polarization scenario. (c) Same as (b) but for TM polarization scenario. In both scenarios, there is a significant enhancement of the received signal in the presence of the metacrystal.} } 
 \label{fig:8}
\end{figure*}

\black{\subsection*{Communication link measurements}} 

\textcolor{black}{To experimentally validate the potential for communication link enhancement enabled by the metacrystal, a separate set of  measurements was performed inside the anechoic chamber, as illustrated in Fig.~\ref{fig:8}(a). 
The metacrystal was positioned to enhance the communication link between the transmitter and receiver in a non-line-of-sight scenario, despite the presence of an obstacle (an absorbing panel) between them.
In this experiment, we validate only the anomalous reflection functionality for normally incident waves.
To maintain a setting closer to real-world conditions, several supporting stands within the anechoic chamber were left uncovered with absorbers, introducing additional sources of scattering. The distance from the TX and RX to the metacrystal central point is about 1.80~m and 5.22~m, respectively.}

\black{For these measurements, we fabricated a metacrystal with the same topology as in the previous experiments, but with larger footprint with dimensions $33.6\lambda \times 42\lambda \times 3.26\lambda$ ($20.16~\mathrm{cm}\times 25.20~\mathrm{cm}\times 1.95~\mathrm{cm}$), as show in the left panel of Fig.~\ref{fig:8}(a). To achieve this large size, we alligned four smaller printed metacrystal sections edge-to-edge along the $y$-direction. 
The tiled build preserves the designed unit-cell response while enabling straightforward scaling to larger apertures, highlighting the metacrystals suitability for mass fabrication.}

\black{Figures~\ref{fig:8}(b) and~\ref{fig:8}(c) show the measured magnitude of the received signal at the receiving antenna (RX) for TE and TM polarizations, respectively, versus the incident angle $\theta_{\rm inc}$, defined as the angle between the normal to the metacrystal plane and the direction to the transmitting antenna (TX) for the  frequency of 51.6~GHz. The solid and dashed lines depict the data with and without metacrystal placed in the communication environment.  
A maximum in the received signal is observed at $\theta_{\rm inc}=0^\circ$, in agreement with the design specifications.
At this angle, the received signal is enhanced by approximately 20~dB and 24~dB for TE and TM polarizations, respectively, compared to the same environment without the metacrystal. These results demonstrate the strong signal reception improvement due to metacrystal for both polarizations.}

\textcolor{black}{To relate the measured reflection response to the communication performance, we estimate the achievable Shannon channel capacity for the given environment according to
$C = B \log_2 \!\left( 1 + \frac{P}{kTB} \right)$, 
where $B$ is the measured bandwidth, $P$ is the received signal power averaged over the bandwidth, and $kT$ denotes the thermal noise power density. 
Here, we define the operational bandwidth of the metacrystal  as the common frequency range where both TM and TE received signals remain within half power range of their respective maxima. 
As shown in Supplementary Section 3, the common operational bandwidth spans from \text{49.1~GHz} to \text{51.7~GHz}. 
In the considered environment, the channel capacity was enhanced by 51\% (TM) and 139\% (TE) relative to the reference cases without the metacrystal. In absolute terms, the channel capacity improvements correspond to 14.3~Gbps and 8.2~Gbps for TM and TE polarizations, respectively.
Thus, the anomalous reflection provided by the metacrystal translates directly into an improved signal-to-noise ratio and, consequently, a higher achievable data throughput within the operational band.}

\section*{Discussion}
We present an approach for the simultaneous manipulation of multiple wave properties, with significant potential to impact a broad range of applications in wireless communications. We demonstrated the effectiveness of this approach through the design, fabrication, and measurement of a multifunctional metacrystal. In the present work, the demonstrated single-nozzle, low-cost FDM fabrication route is directly applicable up to $\sim 100$~GHz, which already covers the most widely discussed near-term 6G-relevant spectrum ranges, including mm-wave spectrum in the 24--71~GHz range~\cite{RohdeSchwarz_PathTowards6G_2024}. Importantly, the inverse-design framework itself is fabrication-agnostic: extending metacrystals to sub-THz and THz frequencies would primarily require higher-resolution manufacturing, with different cost/throughput trade-offs than the low-cost FDM route used here (e.g., two-photon polymerization microfabrication, where feature-size control down to $\sim 100$~nm is available).

In contrast to intelligent surfaces that are typically narrowband and perform single functionality at a time, the metacrystals, although non-tunable, are capable of multiple concurrent and independent wave transformations, with diffraction efficiencies approaching 100\% under lossless material assumptions and the possibility of broadband frequency response. In terms of the angular coverage for beam steering, metacrystals can cover nearly $360^\circ$ angular range, including both reflection and transmission regimes in a single geometry. It is important to note that active reconfigurable intelligent surfaces and passive metacrystal panels address complementary but different deployment regimes: reconfigurable surfaces provide adaptability in highly dynamic environments but require substantial hardware overhead (large number of tunable elements with bias/control electronics), whereas passive metacrystals are much more attractive for static or quasi-static communication environments~\cite{huawei}.

By employing a binarized topology, our approach is fully compatible with conventional 3D-printing manufacturing, which makes it scalable, cost-effective, and suitable for mass production, with the potential for widespread use in wireless communication networks. We estimate that the fabrication cost (consumables) of a metacrystal with a surface area similar to that shown in Fig.~\ref{fig:8}(a) is only 15 USD. This low consumables cost highlights the economic viability of the proposed panels for large-area deployment. It is worth mentioning that in practical installations, the metacrystal panel could be packaged for environmental durability, for example, using a thin low-loss radome or encapsulation layer, and supported by routine maintenance to preserve its long-term performance.
Additionally, the generic nature of our method enables others to build upon this work and explore new opportunities in wireless communication systems.

\section*{Methods}

\subsection*{Topology Optimization}

We use the adjoint-based optimization approach for the design of metacrystals due to its high efficiency and integration with the rigorous coupled-wave analysis (RCWA). Specifically, the adjoint method provides exact analytical gradients with respect to design parameters and requires only two simulations (forward and adjoint) per optimization iteration, making it highly scalable and computationally efficient—particularly important for high-dimensional parameter spaces in multilayer photonic structures. The gRCWA package~\cite{jin2020inverse} we employed further facilitates this by automating the gradient computation via automatic differentiation.

We discretize the metacrystal in the $xz$-plane into deeply subwavelength voxels with different permittivity values within a specific range $\varepsilon \in [\varepsilon_{\rm min}; \varepsilon_{\rm max}]$.
The optimized region in the two-dimensional simulation is the unit cell with dimension $(D, T)$ in the $(x, z)$ directions, respectively. 
The design parameters being optimized are the relative permittivity values of each voxel. The design process aims to minimize the following loss function:
\begin{equation}
{\rm LF} = {\sum\limits_{i,n,k}{\left| {R^{\rm obj}_{i,n,k}} - {R^{\rm act}_{i,n,k}} \right|}}^{2},
\label{lossf}
\end{equation}
where $R$ represents the reflectivity (intensity ratio) into a given diffraction order, $i$ is the index introduced to differentiate between various incidences (e.g., $i=1,2,...,6$ for the scenario shown in Fig.~\ref{fig:fig_gen}, as the metacrystal is optimized for six different incident waves in total), $n$ denotes the diffraction order index, and $k$ indicates the polarization (either TE or TM). The indices ``obj" and ``act" refer to the objective and actual calculated reflectivity values, respectively.

We employ RCWA to calculate the reflection  and subsequently assess the loss function as described in Eq.~(\ref{lossf}). Gradients are derived using the autograd package~\cite{jin2020inverse}. 
Binarized permittivity distributions (as shown, for example, in Fig.~\ref{fig2:fig2}) are achieved through the projection method~\cite{wang2011projection} (see details in Supplementary Section 4), where the density functions are initialized with random coherent noise~\cite{zhao2021perfect}. The optimization typically converges after 70 to 500 iterations. Additional details can be found in~\cite{wang2022design}.

Additionally, we implement a blur function to control the minimum feature size of the structure and to smoothen the permittivity distribution\cite{hughes2019forward}


\subsection*{Device Fabrication}
Applying the bi-level topology optimization, we generated a 2D binary permittivity distribution of the metacrystal. The data were imported into COMSOL Multiphysics. Leveraging MATLAB LiveLink scripts, we then extracted a 3D model and generated the corresponding .stl file. The .stl file was then imported into Ultimaker Cura 5.3.1 for final device preparation. The finalized file was uploaded to the Ultimaker S5, and printer settings were optimized to achieve the best possible print quality. The printing process for the final structure took 34 hours and used 18.17 meters of silver-colored PLA. A 0.25 mm nozzle was used to achieve higher resolution in the xy-plane, which increased the printing time. By selecting a 0.4 mm nozzle and a 0.1 mm layer thickness, the printing time could be reduced to 25 hours. To enhance fabrication tolerance, we apply a spatial convolution kernel during the optimization process, which acts as a geometric low-pass filter. This enforces a minimum feature size and suppresses fine-scale variations, making the design more robust to typical 3D printing imperfections. Combined with the non-resonant, low-index nature of the structure, this leads to stable performance under moderate fabrication variability. Several large images of the fabricated sample can be found in Supplementary Section 5.

\subsection*{Permittivity Characterization} 
We used polyacrylic acid (UltiMaker PLA of silver color) for the fabrication of our metacrystal. Since the permittivity of this PLA had not been previously measured in the millimeter-wave frequency range, we performed its measurements using the method proposed in~\cite{wang2022fast}, which is particularly suited for this frequency range. A uniform test sample printed using Ultimaker S5 3D printer was placed between the flanges of two WR-10 waveguides. This method is advantageous because it does not require the sample cross-section to precisely match the waveguide aperture, which is beneficial given the small dimensions of the aperture. 
For the measurements of the S-matrix, we used a vector network analyzer with frequency extenders. The system was calibrated using the Thru-Reflect-Line (TRL) method.

\subsection*{Far-Field Measurements}
Far-field measurements were conducted in an anechoic chamber using a vector network analyzer connected to both transmitting and receiving horn antennas. The antennas used in this experiment were dual-polarized probes (AYSOL, model ASY-CWG-D-400-UG383). They operate in the frequency range of 40 GHz to 60 GHz and support dual polarization, allowing measurements in both horizontal and vertical polarizations. The antennas shown in Fig.~\ref{fig:exp}(a)--(b) provided a gain of approximately 13.0 dBi at 50 GHz.
The metacrystal was located at a distance of 5.69 m (around $95\lambda$) from the transmitting antenna, where the radiation from the antenna can be approximated as a plane wave.  The distance between the metacrystal and the receiving antenna was fixed to 1.75 m (around $29\lambda$). 
To control the metacrystal orientation, it was attached to a platform rotating around the $y$-axis.  
The reflected signal from the metacrystal, indicated by $|S_{21,{\rm m}}|$, was measured by the receiving antenna for various angles $\theta_{\rm r}$. In the reference experiments, the metacrystal was replaced with a metallic plate of the same dimensions without altering the surrounding environment. We tilted the metallic plate by an angle $\theta_{\rm r}/2$ and measured the reflected wave, $|S_{21,{\rm p}}|$, toward the same angle $\theta_{\rm r}$ (the two antennas had the same positions in both measurements). 
To calculate the efficiency of the anomalous reflection from the metacrystal, we additionally calculate the correction factor $\xi_0$ which essentially denotes the ratio of the far-field scattered fields in a given direction from a metal plate and an ideal metacrystal~\cite{diaz2017generalized,asadchy2017flat}. Importantly, $\xi_0 \neq 1$ since the ideal metacrystal and metal plate have different receiving and scattering effective apertures despite equal physical dimensions. Coefficient $\xi_0$
depends on $\theta_{\rm inc}$ and $\theta_{\rm r}$ and is calculated for both TE and TM in the Supplementary Section 2.
The efficiency of the anomalous reflection of the metacrystal  is given by:
\begin{equation}
\xi_{\rm r} = \frac{1}{\xi_0} \frac{|{S}_{21,\mathrm{m}} |}{|{S}_{21,\mathrm{p}} |}.
\end{equation}

\vspace{2cm}
\section*{Data Availability}
The permittivity datasets generated and analysed during this study are provided in the Supplementary Information and Source Data files. The additional data that support the findings are available from the corresponding authors upon request.

\section*{Code Availability}
The custom numerical codes used for electromagnetic simulations based on RCWA and implemented inverse design  are available from the corresponding authors upon request.


\bibliography{references}

\section*{Acknowledgements}
We acknowledge the computational resources provided by the Aalto Science-IT project. We thank Dr. Xuchen Wang and Mr. Kamil I\c{s}\i k for their assistance with the permittivity measurements and electromagnetic software simulations. We also thank Dr. Francisco S. Cuesta, Mr. Javad Shabanpour, Mr. Mostafa Movahediqomi, and Ms. Shadi Safaei Jazi for their support during the experimental phase.
\section*{Funding}
S.F. discloses support for the research of this work from the U.S. Air Force Office of Scientific Research through a MURI project [Grant No. FA9550-21-1-0244]. V.A. discloses support for the research of this work from the Research Council of Finland [Projects No. 365679 and 371367], the Finnish Foundation for Technology Promotion, and the Research Council of Finland Flagship Programme Photonics Research and Innovation (PREIN) [Decision No. 346529]. The remaining authors declare no relevant funding.
\section*{Author Contributions}
M.M.A. performed the electromagnetic design, fabricated the samples, conducted the experiments and measurements, and carried out data acquisition and analysis. M.M.A., P.B.C., and V.A. conceived the idea. S.S.A.Y. assisted with the second set of experimental measurements. H.W. contributed to the development of the computational tools. M.M.A. analyzed the results and prepared the initial manuscript draft. S.F. and V.A. supervised the research. All authors discussed the results, contributed to manuscript revision, and approved the final version.

\end{document}